\documentclass[fleqn,usenatbib]{mnras}

\usepackage{newtxtext,newtxmath}

\usepackage[T1]{fontenc}

\DeclareRobustCommand{\VAN}[3]{#2}
\let\VANthebibliography\thebibliography
\def\thebibliography{\DeclareRobustCommand{\VAN}[3]{##3}\VANthebibliography}

\usepackage{graphicx}	
\usepackage{amsmath}	

\usepackage{float}

\defcitealias{Prole2022}{LP22}
\defcitealias{Schauer2021}{AS21}

\title[Cosmological Lyman-Werner simulations.]{From dark matter halos to pre-stellar cores: High resolution follow-up of cosmological Lyman-Werner simulations.}

\author[L. R. Prole]{
Lewis R. Prole,$^{1}$\thanks{E-mail: Prolel@cardiff.ac.uk}
Anna T. P. Schauer,$^{2}$
Paul C. Clark,$^{1}$ 
Simon C. O. Glover,$^{3}$ 
Felix D. Priestley,$^{1}$ 
Ralf~S.~Klessen,$^{3,4}$ 
\\
$^{1}$Cardiff University School of Physics and Astronomy\\
$^{2}$Department of Astronomy, The University of Texas at Austin, Austin, TX 78712, USA\\
$^{3}$Universit\"{a}t Heidelberg, Zentrum f\"{u}r Astronomie, Institut f\"{u}r Theoretische Astrophysik, Albert-Ueberle-Stra{\ss}e 2, D-69120 Heidelberg, Germany\\
$^{4}$Universit\"{a}t Heidelberg, Interdisziplin\"{a}res Zentrum f\"{u}r Wissenschaftliches Rechnen, Im Neuenheimer Feld 205, D-69120 Heidelberg, Germany\\ 
}

\date{Accepted XXX. Received YYY; in original form ZZZ}

\pubyear{2020}

\begin{document}

\label{firstpage}
\pagerange{\pageref{firstpage}--\pageref{lastpage}}
\maketitle

\begin{abstract}
Molecular hydrogen allows cooling in primordial gas, facilitating its collapse into Population III stars within primordial halos. Lyman-Werner (LW) radiation from these stars can escape the halo and delay further star formation by destroying H$_2$  in other halos. As cosmological simulations show that increasing the background LW field strength increases the average halo mass required for star formation, we perform follow-up simulations of selected halos to investigate the knock-on effects this has on the Population III IMF. We follow 5 halos for each of the $J_{21}$ = 0, 0.01 and 0.1 LW field strengths, resolving the pre-stellar core density of $10^{-6}$ g cm$^{-3}$ (10$^{18}$ cm$^{-3}$) before inserting sink particles and following the fragmentation behaviour for hundreds of years further. We find that the mass accreted onto sinks by the end of the simulations is proportional to the mass within the $\sim 10^{-2}$ pc molecular core, which is not correlated to the initial mass of the halo. As such, the IMFs for masses above the brown dwarf limit show little dependence on the LW strength, although they do show variance in the number of low-mass clumps formed. As the range of background LW field strengths tested here covers the most likely values from literature, we conclude that the IMF for so-called Pop III.2 stars is not significantly different from the initial population of Pop III.1 stars. The primordial IMF therefore likely remains unchanged until the formation of the next generation of Population II stars.

\end{abstract}




\section{Introduction}
This first stars are able to form because pristine baryonic gas can collapse within the gravitational potential well of dark matter (DM) halos \citep{Couchman1986,Haiman1996a,Tegmark1997}, heating it to $\sim 5000$ K and facilitating the formation of H$_2$ \citep{Bromm2002}. H$_2$ primarily forms via the radiative association reaction forming H$^{-}$
\begin{equation}
\rm H + e^{-} \rightarrow H^{-} + \gamma,
\end{equation}
followed by the fast associative detachment reaction forming  H$_2$
\begin{equation}
\rm H^{-} + H \rightarrow H_{2} + e^{-}.
\label{eq:H2}
\end{equation}
Radiative cooling from the molecular hydrogen renders the gas gravitationally unstable, which allows it to decouple from the DM and collapse to form the first stars, known as Population III (Pop III) stars. The necessity of H$_2$ renders any process of H$_2$ destruction as a mechanism to delay or prevent Pop III star formation.

While so-called Pop III.1 stars form from purely cosmological initial conditions, the radiation they produce affects Pop III.2 stars that form in its presence. As the masses of Pop III stars are predicted to be larger than present-day counterparts (due to the lack of cooling from dust and metals), they are expected to emit large amounts of ionizing radiation (e.g. \citealt{Schaerer2002}). Ionizing photons above the Lyman limit (13.6 eV) create H$\,${\sc ii} regions around the stars up to the boundary of their Str{\"o}mgren spheres \citep{Whalen2004,Kitayama2004,Alvarez2006, Abel2007,Yoshida2007a,Jaura2022} while photons below the Lyman limit are free to escape their Str{\"o}mgren sphere. Lyman-Werner (LW) photons between 11.2 and 13.6 eV can dissociate H$_2$ via the two-step Solomon process \citep{Field1966,Stecher1967}
\begin{equation}
\rm H_2+ \gamma \rightarrow H^*_2 \rightarrow 2H,
\label{eq:solomon}
\end{equation}
where H$^*_2$ represents an electronically excited state of H$_2$. Photons with energy above 0.76~eV can also photodissociate H$^{-}$ via
\begin{equation}
\rm H^{-} + \gamma \rightarrow H + e^{-},
\label{eq:H-}
\end{equation}
reducing the H$^{-}$ abundance and hence the rate at which H$_{2}$ can form via reaction \ref{eq:H2} (e.g. \citealt{Chuzhoy2007}). This stellar feedback provides a potential obstacle for Pop III.2 stars to overcome during formation. Investigations into the effects of these far UV fields typically categorise the field strength by the intensity in the LW band $J_{21}$, in units of 10$^{-21}$~erg~s$^{-1}$~cm$^{-2}$~Hz$^{-1}$~sr$^{-1}$.

Calculations by \cite{Haiman1997} found that before the Str{\"o}mgren spheres of Pop III stars overlap, the UV background below the ionization threshold was able to penetrate large clouds and suppress their H$_2$ abundance. They also found that the flux necessary for H$_2$ photodissociation is several orders of magnitude smaller than the flux needed to reionize the universe. \cite{Haiman2000} showed that this photodissociation of H$_2$ suppresses further Pop III star formation inside small halos and delays reionization until larger halos form.

Collapse is not impossible without sufficient H$_2$ for cooling. \cite{Omukai2001b} showed that if the LW field is sufficient to keep a halo free of molecular hydrogen, the gas can nevertheless collapse via atomic hydrogen line cooling if the halo has a virial temperature $T_{\rm vir} > 8000$~K. The collapse occurs almost isothermally, possibly resulting in the formation of a direct collapse black hole (DCBH) (e.g. \citealt{Bromm2003b,Spaans2006a,Latif2013b}), a possible progenitor of the supermassive black holes observed at high redshifts (e.g.\ \citealt{Mortlock2011,Matsuoka2019a}). However, for a T = 10$^5$ K blackbody spectrum expected from Pop III stars, a field strength of $J_{21} \sim 10^4$ is required to keep the gas atomic during the collapse \citep{Glover2015,Agarwal2015,Agarwal2016,Sugimura2014}, while the average exposure is expected to be $J_{21} < 0.1$ at $z \sim 15$ \citep{Ahn2009,Trenti2009,Wise2012,Agarwal2012,Skinner2020}.             
\cite{Dijkstra2008} showed that only a fraction of $10^{-8} - 10^{-6}$ of DM halos with virial temperatures $> 10^4$ K have a close luminous neighbour within $< 10$ kpc, and are exposed to an LW flux $J_{21} > 10^3$. The occurrence of atomically cooled halos is therefore expected to be rare.

Studies have shown that values of $J_{21}$ orders of magnitude lower than the critical intensity required to completely suppress H$_{2}$ cooling in massive halos can still drastically affect halo collapse. Typically, the critical mass for efficient molecular hydrogen cooling and subsequent star formation increases with increasing $J_{21}$ (e.g. \citealt{Machacek2001, Yoshida2003,OShea2008,Visbal2014a,Schauer2021}). Cosmological simulations by \cite{Yoshida2003} found gas cooling was suppressed for $J_{21}=0.1$, leading them to predict that star formation
would not occur in halos with $T_{\rm vir} < 8000$~K for LW field strengths greater than this. Conversely, \cite{OShea2008} found that for field strengths as high as $J_{21}=1$, H$_2$ cooling leads to collapse despite the depressed core molecular hydrogen fractions. They also noted that higher LW background fluxes lead to higher accretion rates. High resolution cosmological simulations by \cite{Schauer2021} (hereafter \citetalias{Schauer2021}) examined the impact of different values of the LW field strength on a large sample of minihalos. They showed that both M$_\text{av}$, the average minihalo mass required for efficient H$_{2}$ cooling (i.e.\ the mass above which more than 50\% of minihalos of that mass can cool), and M$_{\rm min}$, the minimum minihalo mass required for efficient cooling, 
increased with increasing $J_{21}$. An increase in the critical halo mass for star formation with increasing $J_{21}$ was also found by \cite{Kulkarni2021}, although they find a significant effect only for $J_{21} > 1$. In contrast, cosmological simulations by \cite{Skinner2020} found no relationship between the LW intensity and host halo mass.

The true average $J_{21}$ intensity is expected to vary with redshift. \cite{Hirano2015} followed the formation and evolution of 1540 star-forming gas clouds. They found that in their models, the characteristic mass of  Pop III stars shifted to lower masses with decreasing redshift due to the radiative feedback of previous generations of stars. For $z > 20$, half of the star-forming gas clouds were exposed to intense FUV radiation, with an average exposure of $J_{21} \sim 0.07$. Due to smaller stellar masses and the expanding distance between stars, the FUV background became weaker at lower redshifts. For  $15 < z < 25$, almost all the clouds had nonzero intensity $J_{21} > 0.01$. The average LW intensity in  \cite{Skinner2020} increased stochastically from 10$^{-3}$ at $z \sim 25$ to 10 at $z \sim 10$. For redshifts above $\sim$12, $J_{21}$ remained $>0.1$.

Self-shielding is a process that occurs when large column density of molecular hydrogen protects the inner regions against photodissociation because one photon can only photodissociate one H$_2$ molecule. This self-shielding allows further H$_{2}$ production and H$_{2}$ cooling \citep{Shang2010, Agarwal2014, Regan2014,  Hartwig2015d}. The large nonequilibrium abundance of electrons in gas cooling from above T $>$ 10$^{4}$ K also boosts H$_{2}$ formation \citep{Oh2002}. Early attempts to model self shielding (e.g. \citealt{Shang2010}) multiplied the intensity in the LW band by a self-shielding factor given by \cite{Draine1996}. \cite{Wolcott-Green2011} showed that this method underestimated the numerically calculated self-shielding rate by more than an order of magnitude in low-density regions, by overestimating shielding by a large factor at temperatures above a few hundred kelvin. They modified the method of Draine $\&$ Bertoldi by estimating the shielding factor based on the Sobolev length, using local properties of the gas. This modification was computationally inexpensive and used in many subsequent investigations into the aforementioned critical intensity required to form atomic halos, typically producing values an order of magnitude lower than those using the original Draine $\&$ Bertoldi shielding (e.g.  \citealt{Glover2015,Agarwal2015,Agarwal2016}). \cite{Clark2012} improved on this method further with their introduction of the TreeCol algorithm, which calculates maps of the column density distribution seen by each computational element in a simulation in a computationally efficient fashion with the help of an oct-tree.
\cite{Hartwig2015a} took this approach further by accounting for the relative velocities between different computational 
elements. This Doppler-shifts the spectral lines, reducing the effectiveness of self-shielding (since molecules shifted by more than the linewidth do not
contribute to the effective column density).

In addition to a background LW field, primordial star formation is complicated further by streaming velocities between the the gas and DM. Prior to recombination, baryons were tightly coupled to photons. As DM does not experience Thomson scattering, there should have been a relative velocity between the DM and baryons (e.g. \citealt{Ma1995}). At recombination, the relative velocity was $\sim$ 30 km~s$^{-1}$ and was coherent over several comoving Mpc. Recombination resulted in a drop in the sound speed to $\sim$ 6 km s$^{-1}$ as the gas transitioned from plasma to a neutral state, meaning the relative velocities were highly supersonic. \cite{Tseliakhovich2010} showed that the presence of these large-scale streaming velocities suppresses the abundance of the first bound objects by advecting small-scale perturbations near the baryonic Jeans scale. Moving-mesh calculations by \cite{Greif2011b} found that the additional momentum and energy from the streaming velocities reduces the gas fractions and central densities of halos, increasing the typical virial mass required for efficient cooling by a factor of three. They also noted that the turbulent velocity dispersion increased in the presence of streaming velocities. The simulations of \citetalias{Schauer2021} found that the increase in the average and minimum halo mass from increasing streaming velocities is additive on top of the effect of a LW field, with streaming velocities having the larger impact of the two.

While it was initally believed that Pop III stars formed in isolation \citep{Haiman1996} and were massive \citep{Abel2002,Bromm2002}, more recent studies show that primordial gas fragments to give rise to a larger populations of lower mass stars \citep{Clark2011, Smith2011, Greif2012, Stacy2013, Machida2013, Stacy2014, Susa2019, Wollenberg2020}. In \cite{Prole2022} (hereafter \citetalias{Prole2022}), we used high resolution simulations of idealised, purely hydrodynamical Pop III star formation to show that a number of cores are ejected from the system with masses capable of surviving until the present day. As small-scale primordial magnetic fields do not appear to prevent disc fragmentation \citep{Prole2022a} and accretion of metals onto the surface of these stars during their lifetime is unlikely (e.g. \citealt{Johnson2011, Tanaka2017}), the question is raised about why these stars have not been found within archeological surveys \cite[see e.g.][]{Beers2005, frebel2015, starkenburg2017}. Since most high resolution simulations of Pop III star formation have considered only the Pop III.1 case, one possible explanation could be that Pop III.2 star formation yields a different IMF, i.e.\ that Pop III stars forming in the presence of a LW background have systematically larger masses than those forming in the absence of a background.  

In this paper, we aim to test this hypothesis by producing the most accurate prediction of the Pop III.1 and Pop III.2 initial mass functions (IMF) to date. We investigate how the increase in halo masses due to increasing LW field intensity changes star formation within them, by performing high resolution follow-up simulations of cosmological halos drawn from the simulations of \citetalias{Schauer2021}. 
The structure of our paper is as follows. In Section \ref{sec:method}, we describe the cosmological simulations of \citetalias{Schauer2021}, our selection criteria for the halos chosen for follow-up simulations, the chemical model we use and our use of sink particles. In Section  \ref{sec:ics_discussion} we review the characteristics of the halos as they are taken from \citetalias{Schauer2021}, before presenting the results of the zoom-in simulations in Section \ref{sec:collapse_discussion}, where we probe the density regime of the molecular core. In Section \ref{sec:IMF_discussion}, we compare the fragmentation behaviour once sink particles have formed and present the IMFs at the end of the simulations. We discuss caveats in Section \ref{sec:caveats} before concluding in Section \ref{sec:conclusion}.

\section{Numerical method}
\label{sec:method}
\subsection{{\sc Arepo}}
The simulations presented here were performed with the moving mesh code {\sc Arepo} \citep{Springel2010} with a primordial chemistry set-up. {\sc Arepo} combines the advantages of AMR and smoothed particle hydrodynamics (SPH: \citealt{Monaghan1992}) with a mesh made up of a moving, unstructured, Voronoi tessellation of discrete points. {\sc Arepo} solves hyperbolic conservation laws of ideal hydrodynamics with a finite volume approach, based on a second-order unsplit Godunov scheme with an exact Riemann solver. Automatic and continuous refinement overcome the challenge of structure growth associated with AMR (e.g. \citealt{Heitmann2008}).

\subsection{Cosmological simulations}
The cosmological simulations performed by \citetalias{Schauer2021} assumed a $\Lambda$CDM cosmology with parameters $h=0.6774$, $\Omega_0 = 0.3089$, $\Omega_{\rm b} = 0.04864$, $\Omega_\Lambda = 0.6911$, $n = 0.96$ and
$\sigma_8 = 0.8159$ as derived by the Planck \cite{PlanckCollaboration2020}.  The simulations were initialised at $z=200$ with an initial DM distribution created by MUSIC \citep{Hahn2011} using the transfer functions of \cite{Eisenstein1998} and the gas distribution initially followed the DM. The DM was represented by $1024^{3}$ particles and the gas was initially modelled with $1024^{3}$ grid cells, all contained within a box with side length $1 \: h^{-1} \: {\rm Mpc}$ in comoving units. During the simulation, an additional Jeans refinement criterion was applied: cells were refined whenever necessary so as to ensure that the Jeans length was always resolved with at least 16 cells. This refinement was carried out until the gas reached a threshold density of $\sim 10^{-19}$ g cm$^{-3}$. Above this threshold density, gravitationally bound and collapsing gas was converted into collisionless sink particles, as explained at greater length in \citetalias{Schauer2021}. The simulations were carried out with four different values of the baryonic streaming velocity ($v_{\rm str} = 0, 1, 2$ and 3, in units of $\sigma_{\rm rms}$, the large-scale root mean squared value) and three different values for the LW field strength ($J_{21} = 0$, 0.01 and 0.1). In this study, we make use of the three simulations with $v_{\rm str} = 1\sigma_{\rm rms}$ because this is the most representative value available, as the volume fraction of streaming velocities peaks at 0.8$\sigma_{\rm rms}$ \citepalias{Schauer2021}.

\subsection{Halo selection}
\label{Sims}
Given snapshots at $z=15$ from the three simulations with $v_{\rm str} = 1$ presented in  \citetalias{Schauer2021}, we have selected 5 halos for each value of $J_{21}$. Halo positions and masses were provided by the friends-of-friends (FoF) algorithm as described in that study. The selection criteria for the halos was as follows: 
\begin{itemize}
\item The halos identified by the FoF algorithm were sorted by their mass difference with M$_{\rm av}$, the average minihalo mass above which minihalos become capable of cooling and forming stars. This average mass was defined in \citetalias{Schauer2021}, following \citet{Schauer2019}, to be the minihalo mass above which more than 50\% of minihalos can cool and form stars.\footnote{Note that M$_{\rm av}$ is typically around a factor of three larger than $M_{\rm min}$, the mass of the least massive minihalo with cool gas.} \citetalias{Schauer2021} report M$_{\rm av}$ at a range of redshifts between $z = 22$ and $z = 14$; here, we adopt their values at $z = 15$. By sorting the halos in this way, we ensure that the halos that we eventually select will have masses close to M$_{\rm av}$, i.e.\ that they are representative of the common case at that redshift.

\item We considered a halo if it contained a cell denser than $\rho_{\rm th}=10^{-22}$ g cm$^{-3}$ within a search radius R$_{\rm search} = 1$ kpc in physical units from the halo's central coordinate. Note that checking the H$_2$ abundance is not necessary, as collapse to this density is not possible without a high H$_2$ abundance outside of the atomically cooling halo scenario. 

\item We check that the halo is sufficiently resolved i.e.\ that the 16 cells per Jeans length criteria has not decreased as the cell sizes approach the maximum resolution of the cosmological simulations. This process begins when the density reaches $\sim 10^{-20}$ g cm$^{-3}$, so we reject halos including a cell above this density. Note that this also ensures that a sink particle has not yet been formed within R$_{\rm search}$. 

\item Lastly, we removed the halo from consideration if there were other dense objects in the vicinity by checking there were no cells above $\rho_{\rm th}$ within the radius R$_{\rm search}/2$ $<$ r $<$ R$_{\rm search}$.  
\end{itemize}

\begin{figure*}
	 \hbox{\hspace{0cm} \includegraphics[scale=0.6]{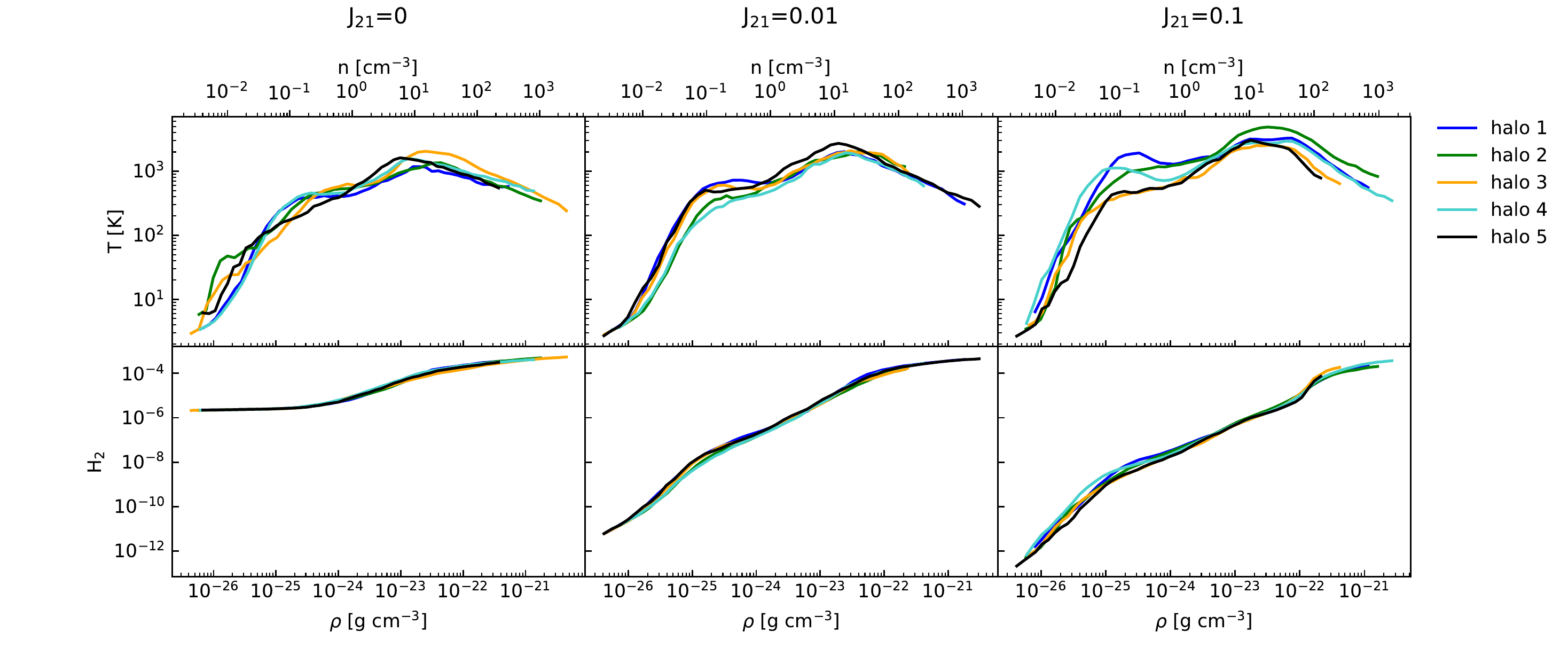}}
    \caption{Summaries of the initial conditions for our simulations. Mass-weighted density profiles of temperature and H$_2$ abundances. }
    \label{fig:ics_radial}
\end{figure*}

\begin{figure*}
	 \hbox{\hspace{-0.1cm} \includegraphics[scale=0.6]{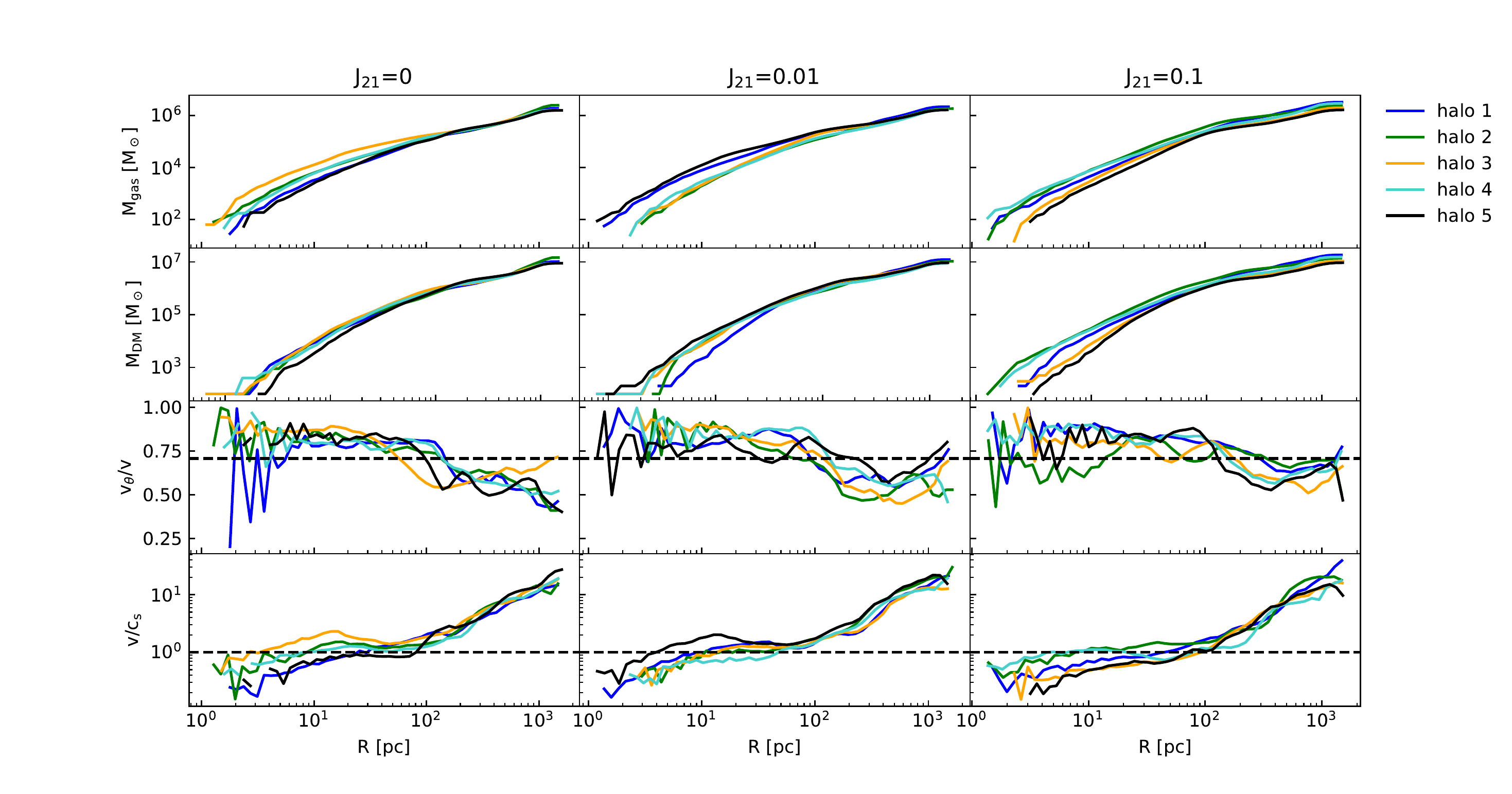}}
    \caption{Summaries of the initial conditions for our simulations. Cumulative radial profiles of gas and DM mass and mass-weighted radial profiles of the ratio of rotational to total velocity (note the dotted line represents the value above which rotational component dominates the velocity) and ratio of velocity to sound speed i.e. Mach number.}
    \label{fig:ics_radial2}
\end{figure*}

\begin{figure*}
	 \hbox{\hspace{-2cm} \includegraphics[scale=0.65]{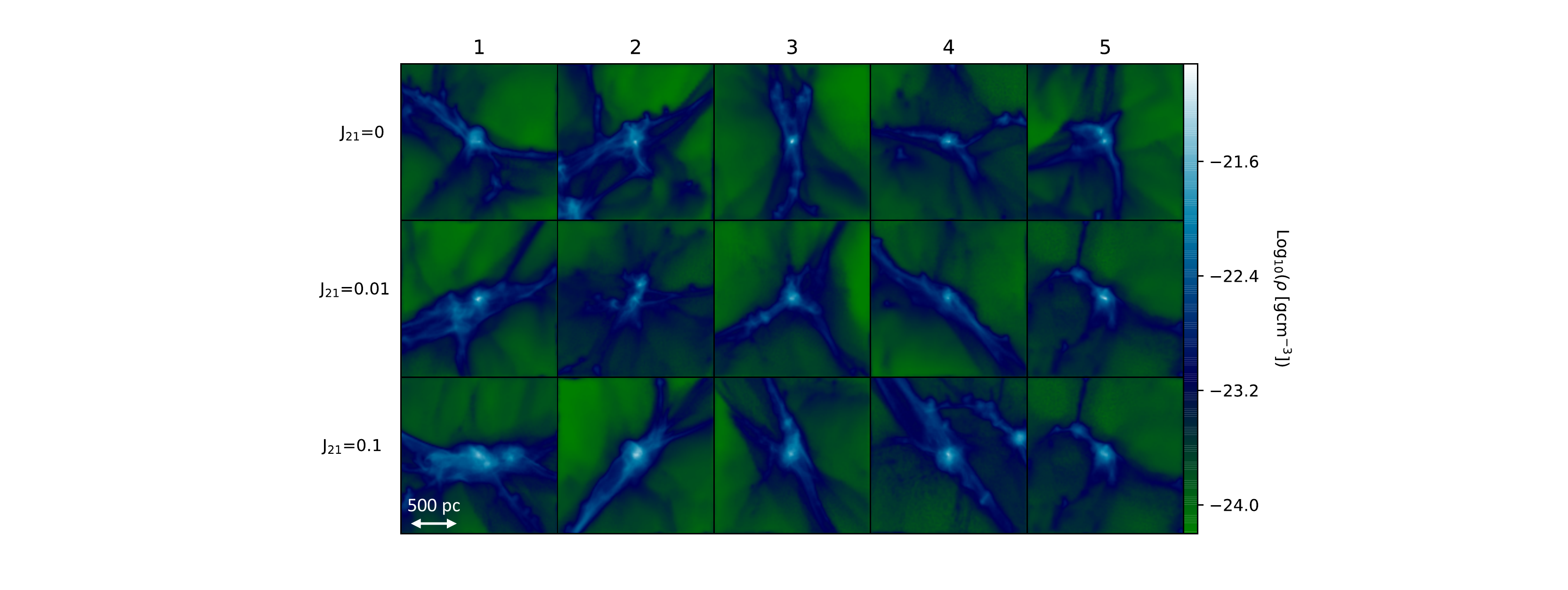}}
    \caption{Column-weighted density projections of the 2~kpc region around the 5 cosmological halos for each $J_{21}$ value, taken from \citet{Schauer2021}, which serve as the initial conditions of the high resolution follow-up simulations presented in this work. The halo numbers to compare with upcoming plots are indicated at the top of the figure.}
    \label{fig:rho_proj}
\end{figure*}

\begin{figure*}
	 \hbox{\hspace{-2cm} \includegraphics[scale=0.65]{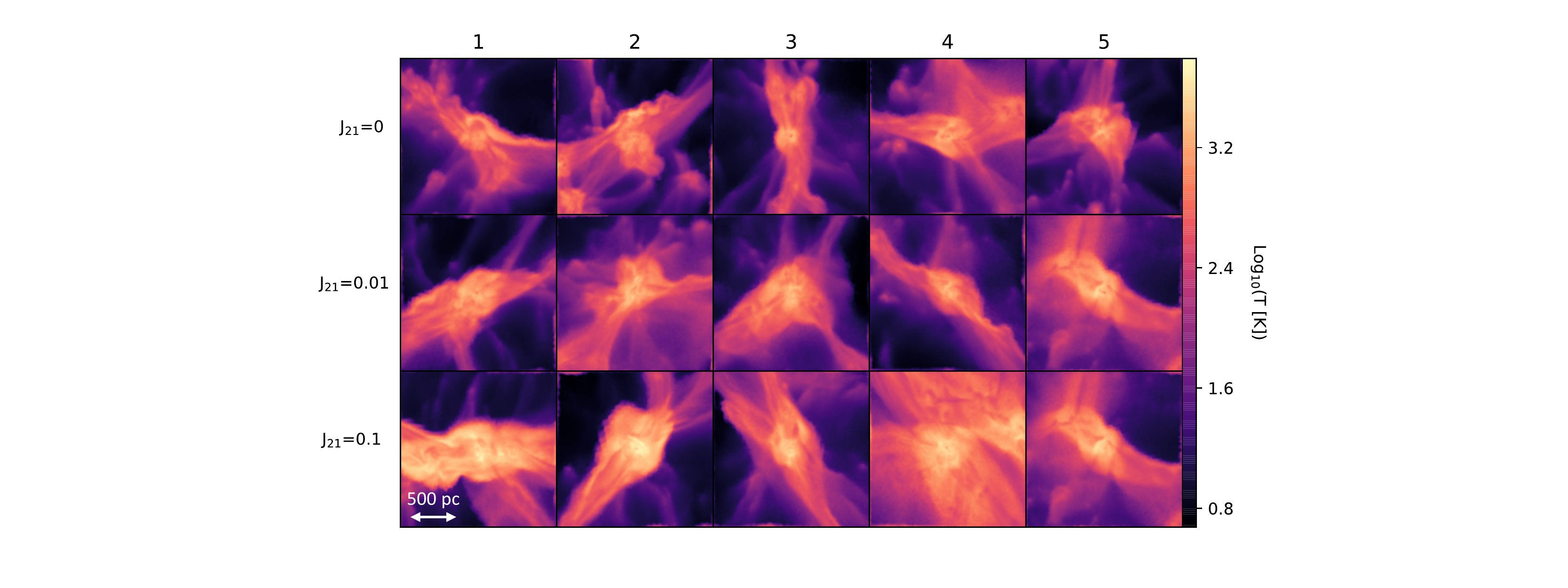}}
    \caption{Column-weighted temperature projections of the 2~kpc region around the 5 cosmological halos for each $J_{21}$ value, taken from \citet{Schauer2021}, which serve as the initial conditions of the high resolution follow-up simulations presented in this work.}
    \label{fig:temp_proj}
\end{figure*}

\begin{figure*}
	 \hbox{\hspace{-2cm} \includegraphics[scale=0.65]{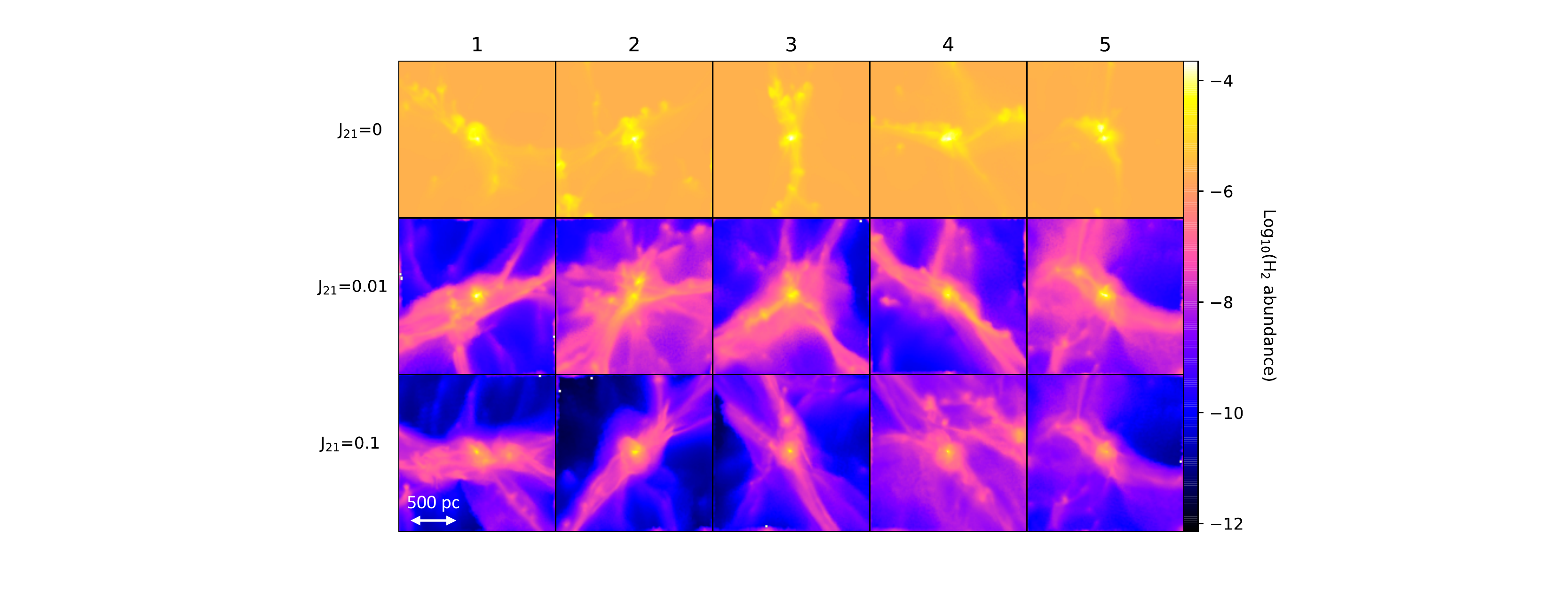}}
    \caption{Column-weighted H$_2$ abundance projections  of the 2~kpc region around the 5 cosmological halos for each $J_{21}$ value, taken from \citet{Schauer2021}, which serve as the initial conditions of the high resolution follow-up simulations presented in this work.}
    \label{fig:H2_proj}
\end{figure*}

We performed this search throughout the simulation cube until 5 suitable halos were selected for each $J_{21}$ value. After converting the units from co-moving to physical, we cropped a 2 kpc box around each of the halos for our zoom-in simulations. The selected halo masses from each simulation are shown in Table \ref{table:1}. Mass-weighted mean temperatures and H$_{2}$ abundances are plotted as a function of density for each halo in Figure \ref{fig:ics_radial} and radial distributions of enclosed gas and DM along with velocity information are shown in Figure \ref{fig:ics_radial2}. Projections of the initial density, temperature and H$_2$ abundance distributions are visualised in Figures \ref{fig:rho_proj}, \ref{fig:temp_proj} and \ref{fig:H2_proj}. Note that all of these figures show the state of the halos at the point at which we extract them from the AS21 simulations, i.e.\ prior to our zoom-in calculation. Our choice of selection criteria means that the same halos are not chosen for direct comparison between each simulation. Rather, they are chosen to be most representative of the Universe at the given background LW field strength. Direct comparison also has the drawback that individual halos will collapse at different redshifts depending on the LW field strength, whereas we exclusively select halos that become able to form stars at $z=15$.

\begin{table}
	\centering
	\caption{Total mass (DM and gas) as detected by the FoF algorithm within the selected halos from \citet{Schauer2021}.}
	\label{table:1}
	\begin{tabular}{lcccr} 
		\hline
		 & \vline & $J_{21}$=0 &  $J_{21}$=0.01 & $J_{21}$=0.1\\
		\hline
		halo & \vline & & M$_{\rm tot}$ [$10^{6}$ M$_\odot$]& \\
		\hline
		 1  &\vline & 3.50 & 4.76 & 8.70\\
	2  &\vline &  3.65 & 4.84 & 8.30\\
	     3  &\vline &  2.81 & 4.30 & 6.70\\
	    4  &\vline &  2.76 & 4.02 & 4.15\\
		 5  &\vline &  2.96 & 3.91 & 3.90\\
		\hline
	\end{tabular}
\end{table}

\subsection{Chemistry}
\label{chem}
Collapse of primordial gas is closely linked to the chemistry involved (e.g.\ \citealt{Glover2006,Yoshida2007, Glover2008, Turk2011}). We therefore use a fully time-dependent chemical network to model the gas. We use the same chemistry and cooling as \cite{Wollenberg2020}, which is described in the appendix of \cite{Clark2011}, but with updated rate coefficients, as summarised in \cite{Schauer2019}. The network has 45 chemical reactions to model primordial gas made up of 12 species: H, H$^{+}$, H$^{-}$, H$^{+}_{2}$ , H$_{2}$, He, He$^{+}$, He$^{++}$, D, D$^{+}$, HD and free electrons. Optically thin H$_{2}$ cooling is modelled as described in \citet{Glover2008}: we first calculate the rates in the low density ($n \rightarrow 0$) and LTE limits, and the smoothly interpolate between them as a function of $n / n_{\rm cr}$, where $n_{\rm cr}$ is the H$_{2}$ critical density. To compute the H$_{2}$ cooling rate in the low density limit, we account for the collisions with H, H$_{2}$, He, H$^{+}$ and electrons. To calculate the H$_{2}$ cooling rate in the optically thick limit, we use an approach introduced by \citet{Yoshida2006}, as described in detail in \citet{Clark2011}. Prior to the simulation, we compute a grid of optically thick H$_{2}$ cooling rates as a function of the gas temperature and H$_{2}$ column density. During the simulation, if the gas is dense enough for the H$_{2}$ cooling to potentially be in the optically thick regime ($\rho > 2 \times 10^{-16} \: {\rm g \: cm^{-3}}$), we interpolate the H$_2$ cooling rate from this table, using the local gas temperature and an estimate of the effective H$_{2}$ column density computed using the Sobolev approximation. In addition to H$_{2}$ cooling, we also account for several other heating and cooling processes: collisionally-induced H$_{2}$ emission, HD cooling, ionisation and recombination, heating and cooling from changes in the chemical make-up of the gas and from shocks, compression and expansion of the gas, three-body H$_{2}$ formation and heating from accretion luminosity. For reasons of computational efficiency, the network switches off tracking of deuterium chemistry\footnote{Note that HD cooling continues to be included in the model.} at densities above 10$^{-16}$~g~cm$^{-3}$, instead assuming that the ratio of HD to H$_{2}$ at these densities is given by the cosmological D to H ratio of 2.6$\times$10$^{-5}$. The adiabatic index of the gas is computed as a function of chemical composition and temperature with the {\sc Arepo} HLLD Riemann solver.

As in \cite{Schauer2021}, we use a radiation field expected from massive Pop III stars. We use a blackbody spectrum at temperature $10^5$ K for energies below 13.6~eV. Above this energy, the flux is expected to drop due to absorption in the intergalactic medium, so we set the value of the radiation field to 0 above 13.6~eV. We model the effects of H$_2$ self-shielding using the TreeCol algorithm \citep{Clark2012}.

\subsection{Sink particles}
If the local Jeans length falls below the minimum cell size of the mesh, artificial collapse occurs. We insert sink particles into the simulations at a threshold density to prevent artificial collapse when the simulation reaches its maximum refinement level. Our sink particle implementation was introduced in \cite{Wollenberg2020} and \citet{Tress2020}. A cell is converted into a sink particle if it satisfies three criteria: 1) it reaches a threshold density; 2) it is sufficiently far away from pre-existing sink particles so that their accretion radii do not overlap; 3) the gas occupying the region inside the sink is gravitationally bound and collapsing. Likewise, for the sink particle to accrete mass from surrounding cells it must meet two criteria: 1) the cell lies within the accretion radius; 2) it is gravitationally bound to the sink particle. A sink particle can accrete up to 90$\%$ of a cell's mass, above which the cell is removed and the total cell mass is transferred to the sink.

Increasing the threshold density for sink particle creation drastically increases the degree of fragmentation, reducing the masses of subsequent secondary protostars (\citetalias{Prole2022}). Ideally, sink particles would be introduced when the gas becomes adiabatic at $\sim 10^{-4}$ g cm$^{-3}$. This is currently computationally challenging. However, the zero metallicity protostellar model of \cite{Machida2015} suggests that stellar feedback kicks in to halt collapse at $\sim 10^{-6}$ g cm$^{-3}$ (10$^{18}$ cm$^{-3}$), so we choose this as our sink particle creation density.

The initial accretion radius of a sink particle $R_{\text{sink}}$ is chosen to be the Jeans length $\lambda_{\text{J}}$ corresponding to the sink particle creation density and corresponding temperature. At $10^{-6}$ g cm$^{-3}$, we take the temperature value from \citetalias{Prole2022} of 4460 K to give a Jeans length of $1.67 \times 10^{12}$ cm. We set the minimum cell length to fit 8 cells across the sink particle in compliance with the Truelove condition, by setting a minimum cell volume $V_{\text{min}}=(R_{\text{sink}}/4)^3$. The minimum gravitational softening length for cells and sink particles $L_{\text{soft}}$ is set to $R_{\text{sink}}/4$. 

The increasing radius of a Pop III protostar is dependent on both its mass and accretion rate (e.g. \citealt{ Omukai2003,Hosokawa2009, Hosokawa2012, Hirano2014}). We allow the sink particle accretion radius $R_{\text{sink}}$ to vary throughout its accretion history, using on-the-fly calculations of the stellar radius using an approximate analytic formulae originally derrived by \cite{Stahler1986}:
\begin{equation}
R_{\text{sink}}=26R_\odot \left(\frac{M}{M_\odot}\right)^{0.27} \left(\frac{\dot M}{10^{-3}M_\odot \text{yr}^{-1}}\right)^{0.41},
\label{eq:radius}
\end{equation}
where we smooth $\dot M$ by taking the average over the time taken to accrete 0.1M$_\odot$. 

The sink particle treatment also includes the accretion luminosity feedback from \cite{Smith2011}, as implemented in {\sc Arepo} by \citet{Wollenberg2020}. Stellar internal luminosity is not included in this work because the Kelvin-Helmholtz times of the protostars formed in our simulations are much longer than the period simulated, meaning that none will have yet begun nuclear burning. We also include the treatment of sink particle mergers used in \citetalias{Prole2022}.

\section{Initial halo characteristics}
\label{sec:ics_discussion}
Figures \ref{fig:ics_radial} and \ref{fig:ics_radial2} show some of the characteristics of the halos at the time when they were selected for zoom-in follow-up simulations. The top panel of Figure \ref{fig:ics_radial} shows that the gas has already been shock heated as it fell into the gravitational potential well of the DM halo, allowing it to produce the necessary H$_2$ to cool and collapse to higher densities. The bottom panel shows the destructive impact of the LW radiation field on the H$_2$ abundance in the outer regions of the halo. The H$_2$ abundance in the central regions reaches the same peak value of $4\times 10^{-4}$ due to self-shielding from the radiation field, however this happens at increasingly higher densities for larger $J_{21}$ values. Figure \ref{fig:ics_radial2} shows that within $\sim 100$ pc, the gas velocity is dominated by the component perpendicular to the radius vector, $v_{\rm \theta}$. The gas surrounding the halos is highly supersonic due to large-scale streaming, which cascades down to become subsonic at scales smaller than $\sim 10$ pc, although we show in Section \ref{sec:collapse_discussion} that the velocities do not remain subsonic once the gas collapses further. From the density projections of Figure \ref{fig:rho_proj}, halo sizes range from $\sim 100-200$~pc, their shapes range from roughly spherical to structurally complex, and each is embedded within a network of filamentary structures. Figure \ref{fig:temp_proj} shows that these webs of filaments are a few hundred kelvin hotter than the $\sim 10$ K gas that surrounds them, with the central halo reaching $\sim 1000-2000$ K. Figure \ref{fig:H2_proj} shows how the background H$_2$ abundance is reduced drastically by the LW background, with significant levels of H$_2$ only appearing in the central regions of the halo.

\section{Further collapse}
\label{sec:collapse_discussion}
\begin{figure*}
	 \hbox{\hspace{-0cm} \includegraphics[scale=0.6]{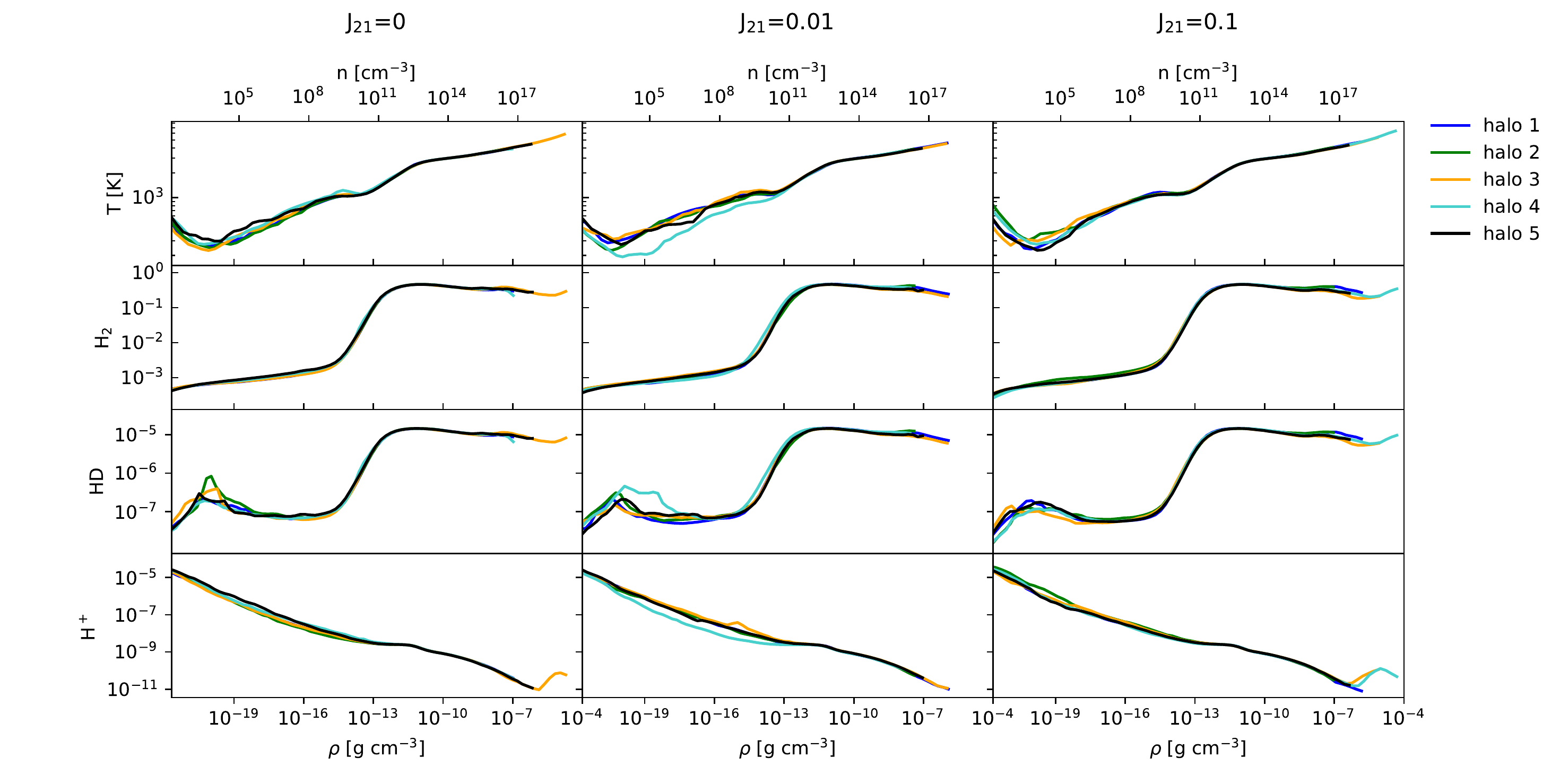}}
    \caption{Mass-weighted temperature, H$_2$, HD and H$^+$ abundances versus density at a time shortly after the formation of the first sink particle. (Note that the HD abundance is only tracked self-consistently up to $\rho = 10^{-16} \: {\rm g \: cm^{-3}}$; above this, we simply fix the HD/H$_{2}$ ratio at 2.6$\times$10$^{-5}$, as explained in Section~\ref{chem}).  
    }
    \label{fig:sink_radial}
\end{figure*}

\begin{figure*}
	 \hbox{\hspace{-0.4cm} \includegraphics[scale=0.6]{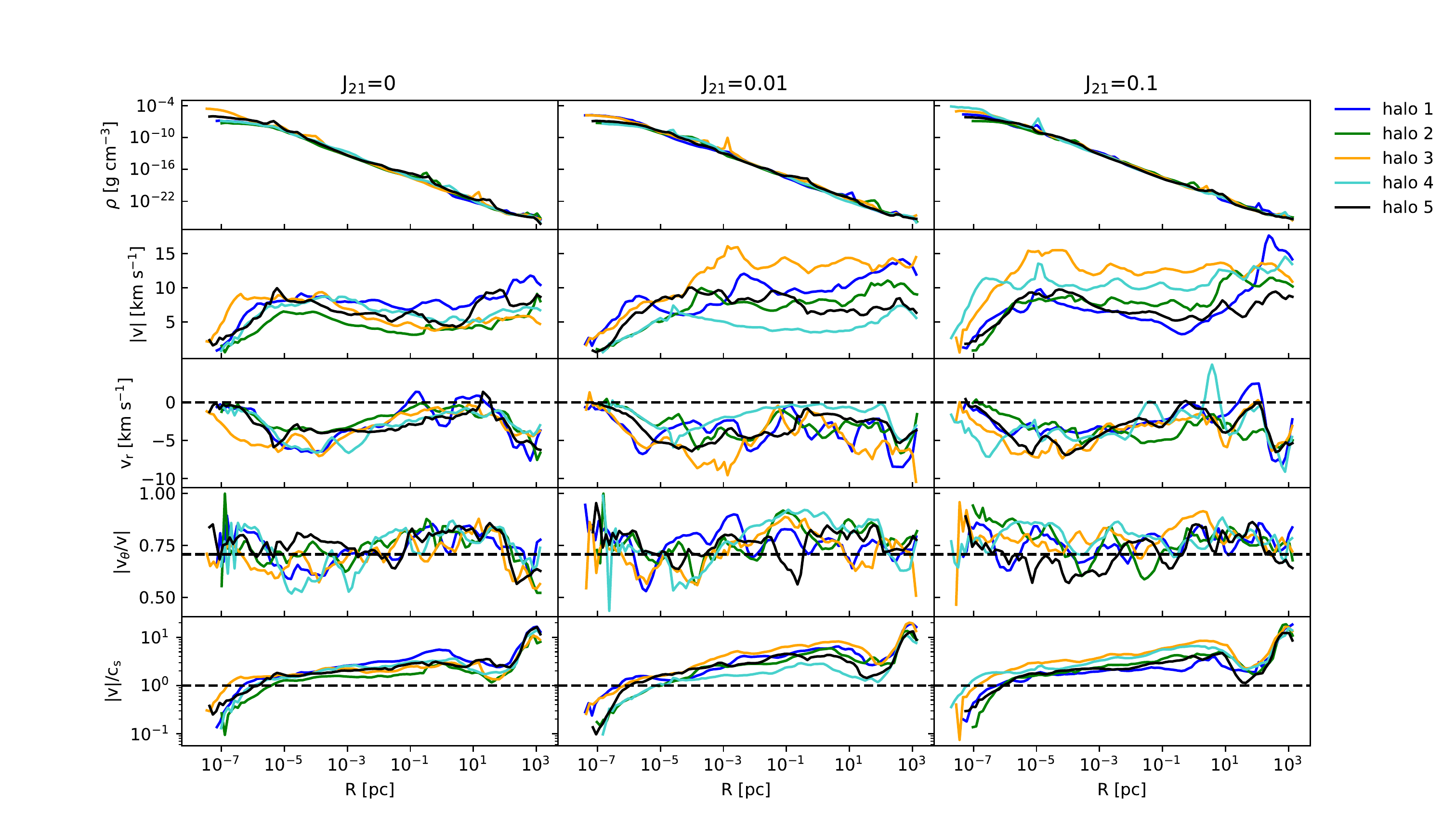}}
    \caption{Radial distribution of cumulative gas mass and mass-weighted radial profiles of radial velocity, ratio of rotational to total velocity (note the dotted line represents the value above which rotational component dominates the velocity) and ratio of velocity to sound speed, taken at a time shortly after the formation of the first sink particle. }
    \label{fig:sink_radial2}
\end{figure*}

We continue the collapse down to densities of 10$^{-6}$ g cm$^{-3}$ before inserting sink particles. Figure \ref{fig:sink_radial} shows temperature and chemical abundance profiles as a function of density, just before the formation of the first sink particle. At densities above 10$^{-15}$ g cm$^{-3}$, three-body H$_2$ formation raises its abundance to over 0.1 within the \emph{molecular core}. The abundance of free electrons falls off with density as the gas recombines. Figure \ref{fig:sink_radial2} shows radial profiles of density and velocity. The rotational component of the gas velocity remains dominated by rotation only down to scales of $\sim 1$ pc, below which  infall begins to dominate. The velocities remain supersonic down to scales of $10^{-6}$ pc (0.2 au). Halos experiencing a LW field are capable of achieving higher velocities, likely due to the higher halo mass. 

\begin{figure*}
	 \hbox{\hspace{-1cm} \includegraphics[scale=0.6]{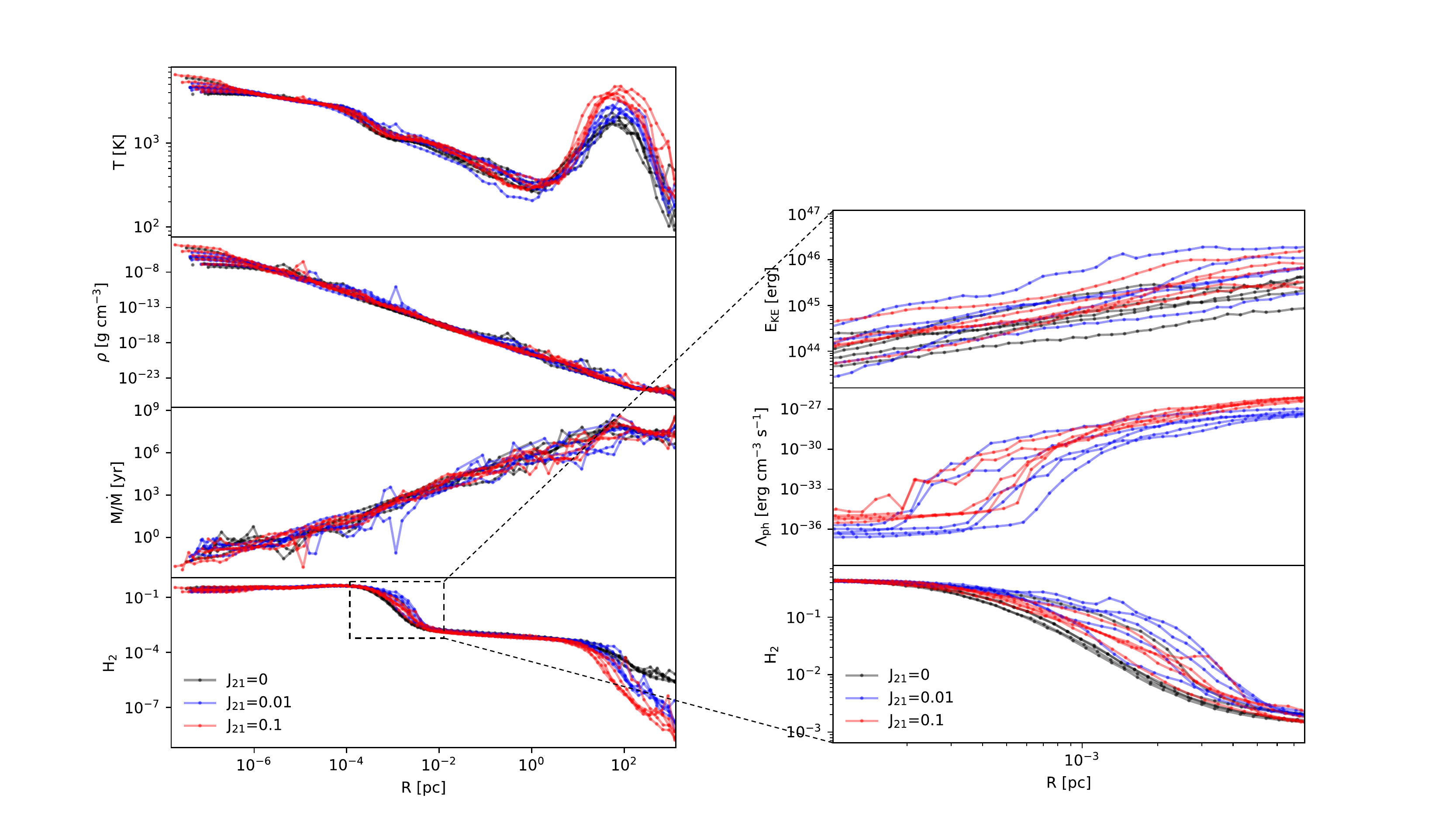}}
    \caption{Left: Comparison of the mass-weighted temperature, density and  and H$_2$ abundance profiles between the $J_{21}$ values. Right: Zoom-in of the transition to fully molecular core, showing the total kinetic energy within  radial shells, mass weighted H$_2$ photodissociation heating rate and H$_2$ abundance.}
    \label{fig:overplot}
\end{figure*}

The left hand side of Figure \ref{fig:overplot} compares the temperature, density, accretion timescale and H$_2$ abundance radial profiles for the different $J_{21}$ values just before the formation of the first sink particle. Stronger LW fields require higher mass halos for star formation, as their stronger gravitational potential is capable of shock-heating the gas to higher temperatures, which increases the H$_2$ formation rate enough to build up a high column density of H$_2$ in order to self-shield the collapsing regions. Despite larger halo masses and shock-heating to higher initial temperatures, the density profile of the gas remains unaffected. Following
\citet{Abel2002} and \citet{OShea2008}, we have also estimated the accretion timescale ($t_{\rm acc} = {\rm M} / \dot{\rm M}$) where
\begin{equation}
\dot{\rm M}= 4 \pi R^2 \rho(R) v_{\rm rad}(R)
\end{equation}
is our estimate of the mass inflow rate at radius $R$ and $\rho(R)$ and $v_{\rm rad}(R)$ are the mass-weighted density and radial velocity within shells at radius $R$. For $t_{\rm acc} < 10^{4} \: {\rm yr}$, there is very little difference between the runs, suggesting that the accretion rate at early times is not influenced by the LW field. For larger $t_{\rm acc}$, we do see a difference between different runs, but this manifests as an increased scatter in $t_{\rm acc}$ at a given $R$ rather than any systematic dependence on the LW field strength. 

The right hand side of Figure \ref{fig:overplot} shows the gas as it transitions into a fully molecular state within the inner $\sim 10^{-2}$ pc core corresponding to the density regime above 10$^{-15}$ g cm$^{-3}$. Halos illuminated gy a LW background have higher gas kinetic energies owing to their larger masses, which promotes a larger molecular core. However, the increasing photodissociation rate with increasing $J_{21}$ acts against this mechanism, reducing the H$_2$ formation rate and shrinking the molecular core. This results in the $J_{21}=0.01$ halos having molecular cores that extend to larger radii than the $J_{21}=0.1$ halos, despite both having larger molecular cores than the $J_{21}=0$ halos. As we only have access to these 3 values of $J_{21}$, the value where the core size is maximised could lie anywhere between $0<J_{21}<0.1$.

\begin{figure*}
	 \hbox{\hspace{-1.2cm} \includegraphics[scale=0.65]{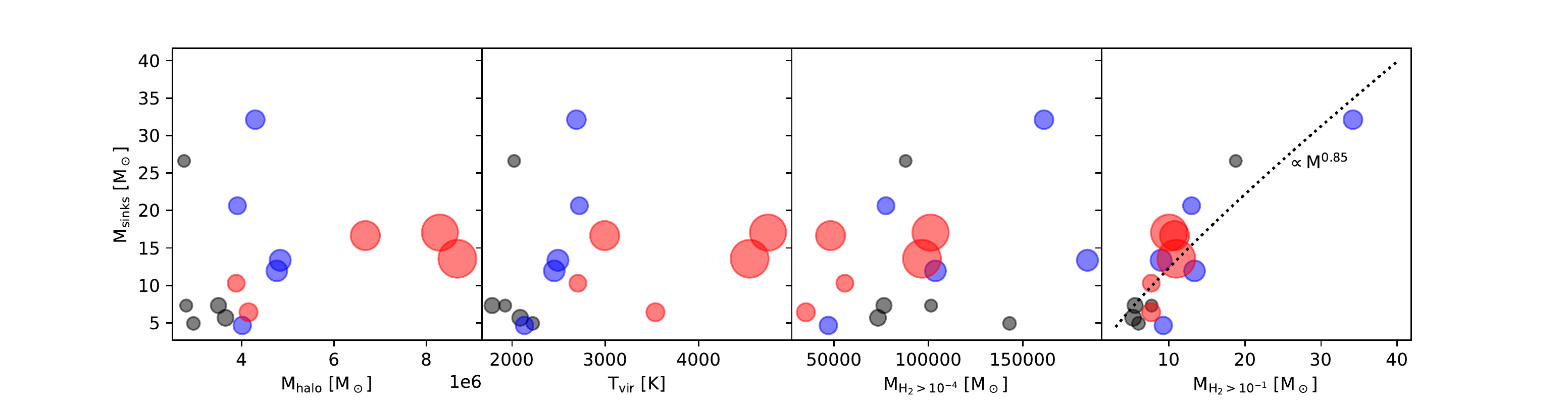}}
    \caption{Total mass in sinks at the end of the simulations versus halo mass, virial temperature, mass within the halo with H$_2$ abundance $>10^{-4}$ and mass within the molecular core with H$_2$ abundance $>10^{-1}$. The size of the markers is scaled with the halo mass.}
    \label{fig:H2core}
\end{figure*}

The importance of the molecular core is shown in Figure \ref{fig:H2core}, which shows the total mass in sink particles at the end of the simulations as functions of the halo mass, virial temperature and mass within different regimes of the collapse, just before the maximum density was reached. The mass in sink particles grows almost linearly with the mass within the inner molecular core with H$_2$ abundances above $10^{-1}$ as 
\begin{equation}
\rm log_{10}(M_{sinks})=(0.85 \pm 0.11) log_{10}(M_{\rm H_2 > 10^{-1}}) + (0.14 \pm 0.24).
\end{equation}
Due to competing effects between halo mass and H$_2$ photodissociation rate, the mass in sink particles is not correlated with the total halo mass or the subsequent mass that initially falls into its potential well with H$_2$ abundances $>10^{-4}$. However, we have only followed the accretion for 300 yr, which corresponds to a free-fall time for gas at density 10$^{-14}$ g cm$^{-3}$. Gas below this density will not have been accreted within the simulation time, while gas above this density resides within the molecular core (see Figure \ref{fig:sink_radial}). It is therefore unclear if the relationship between accretion and mass of the molecular core would remain if the simulations ran for a longer time. If the relationship does hold, we speculate that increasing the size of the streaming velocities between gas and DM may have a greater effect on the IMF, since this also increases halo masses without the counteracting effects of H$_2$ dissociation (\citealt{Tseliakhovich2010,Greif2011b,Hirano2017b,Schauer2019}, \citetalias{Schauer2021}).

\section{Fragmentation and the IMF}
\label{sec:IMF_discussion}
\begin{figure*}
	 \hbox{\hspace{-3.2cm} \includegraphics[scale=0.7]{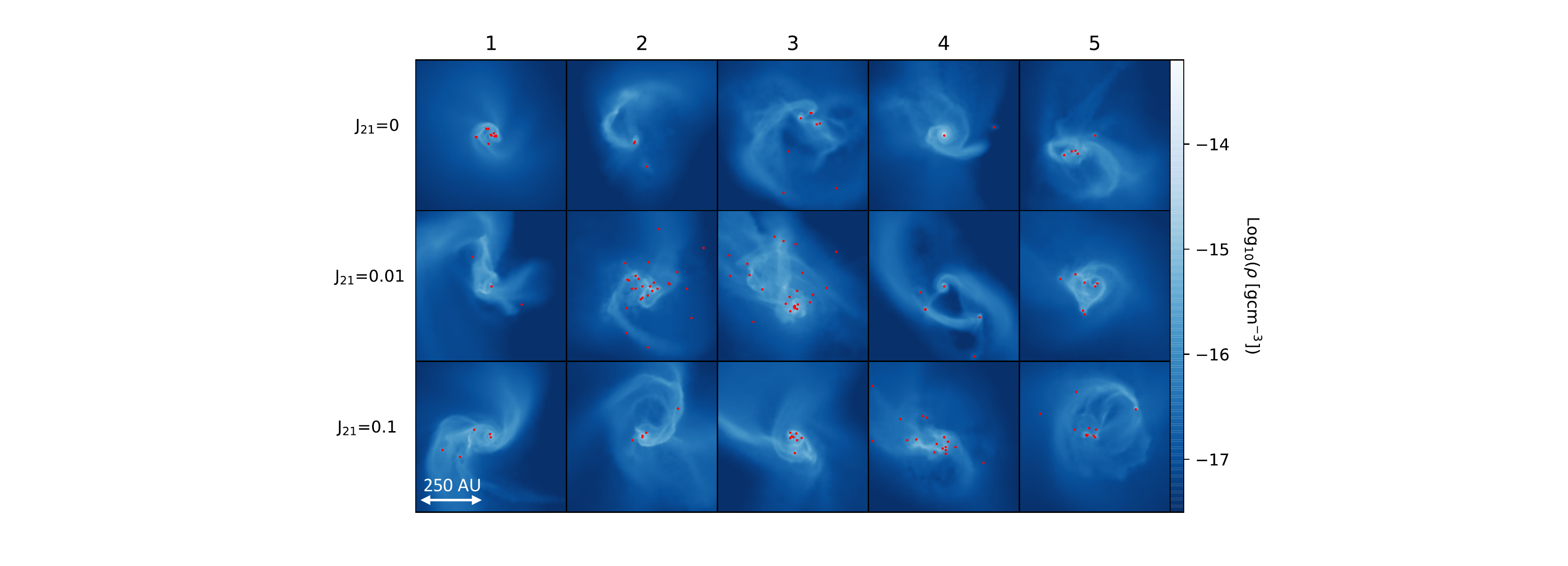}}
    \caption{Column-weighted density projections of the inner 650 AU of the halos at a time 300 yr after the formation of the first sink particle. Sink particles are represented as red dots.}
    \label{fig:sinks}
\end{figure*}

\begin{figure*}
	 \hbox{\hspace{1cm} \includegraphics[scale=0.6]{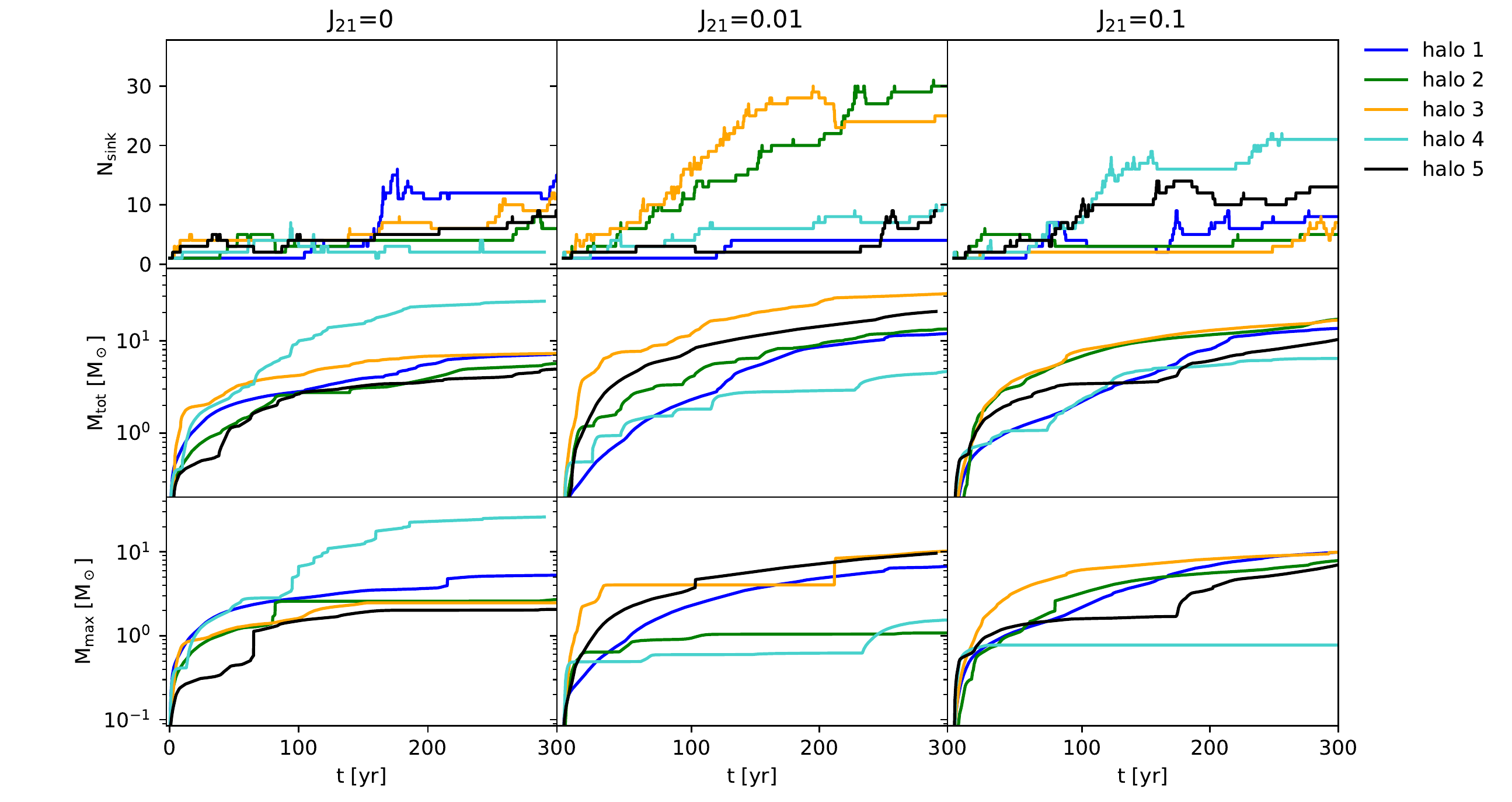}}
    \caption{Number of sink particles formed, total mass in sinks, largest mass sink particle and median mass of sink particles as a function of time.}
    \label{fig:frag}
\end{figure*}

The IMF of Pop III stars is determined by the fragmentation behaviour of the disc around the initial central object and the subsequent accretion onto fragments. Density projections of the inner 650 AU of the halos are shown in Figure \ref{fig:sinks}, while Figure \ref{fig:frag} shows the evolution of the total number of sink particles formed, the total mass in sink particles and highest mass sink particle in each halo as a function of time. While the halos from the simulation without a LW background typically yield less fragmentation, the overall fragmentation behaviour is stochastic, as expected. The total mass accreted onto sinks is typically higher in the halos illuminated by a LW field due to their higher halo mass. 

\begin{figure}
	 \hbox{\hspace{-1cm} \includegraphics[scale=0.53]{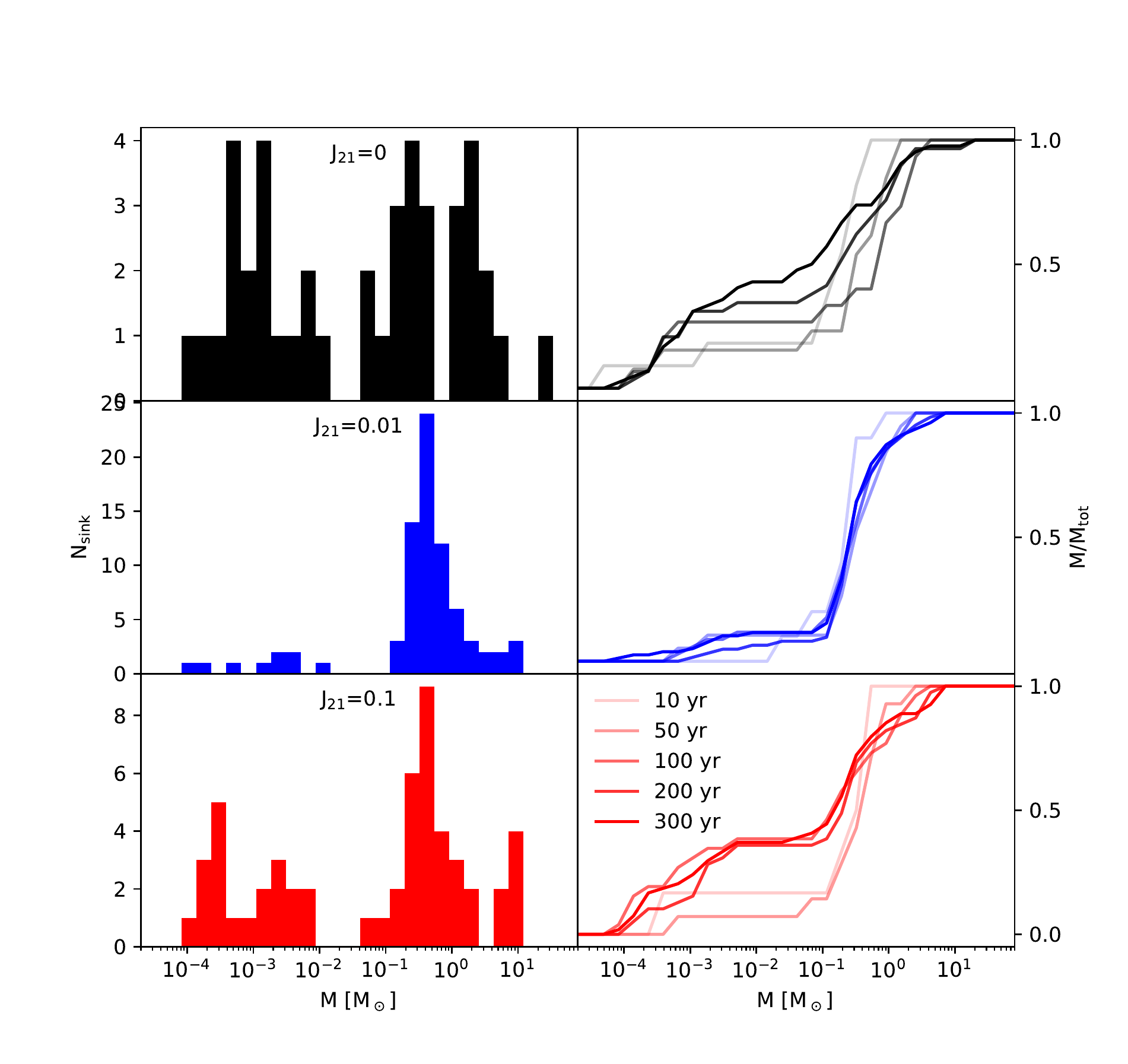}}
    \caption{Left: combined IMFs of sink particles for the different the $J_{21}$ values, Right: Cumulative IMFs of the combined sink particles for the  different the $J_{21}$ values.}
    \label{fig:IMF}
\end{figure}

\begin{figure*}
	 \hbox{\hspace{-1.5cm} \includegraphics[scale=0.7]{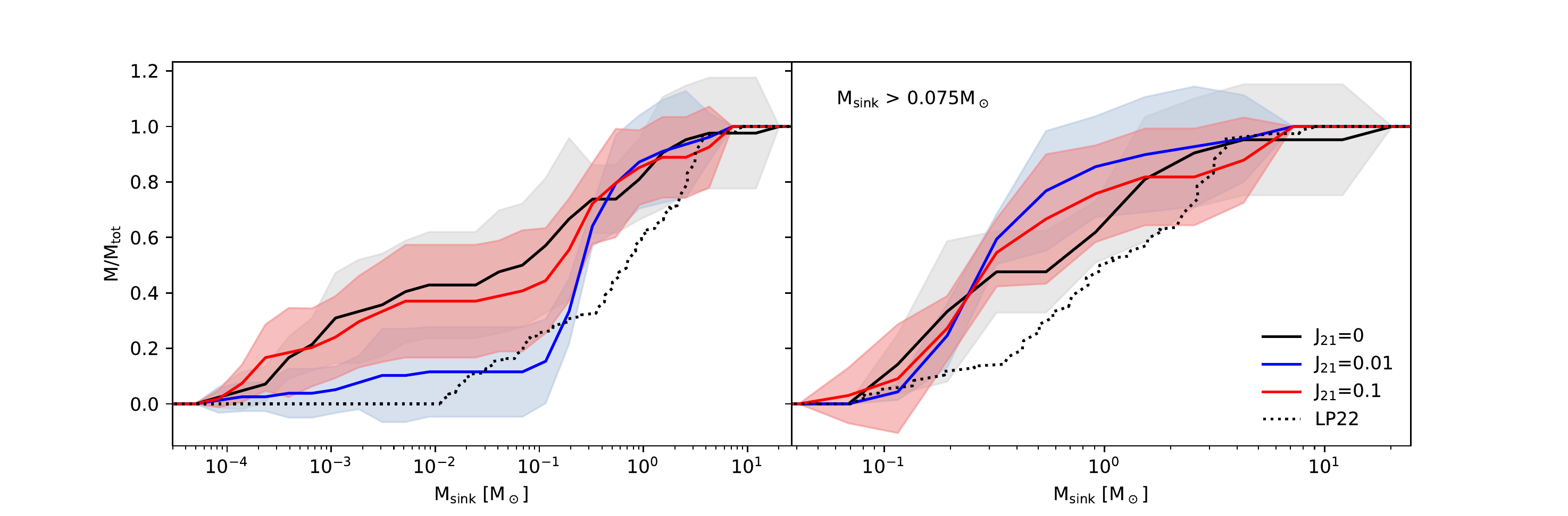}}
    \caption{Left: Comparison of the cumulative IMFs at 300 yr after the formation of the first sink particle. Right: Cumulative IMFs ignoring brown dwarfs ($< 0.075$M$_\odot$).}
    \label{fig:IMF2}
\end{figure*}

Figure \ref{fig:IMF} shows the IMF at a time 300 yr after the formation of the first sink particle. The peak of the IMF positioned at $\sim$0.2-0.5~M$_\odot$ shows little dependence on the LW strength, likely because the positive influence of larger halo masses on the molecular core is regulated by the increasing photodissociation rate. We also show the evolution of the cumulative IMFs in time, which converge by the end of the simulations for the $J_{21}$=0.01 and 0.1 suites. The left side of Figure \ref{fig:IMF2} compares the combined cumulative IMFs for different $J_{21}$ values. While the high mass end of the IMFs are nearly identical between the different $J_{21}$ values, the low mass end of the IMFs show variance due to the random and stochastic nature of ejection events for low mass objects. Assuming these low mass objects ($<0.075$M$_\odot$) do not go on to accrete significant mass, they will remain as brown dwarfs and never sustain nuclear fusion. The right side of the plot shows the cumulative IMFs if the brown dwarfs are ignored. Here, the IMFs fit well within each other's regions of uncertainty, which are the standard deviations from the cumulative IMFs of the 5 halos individually. The overlap between the regions of uncertainty indicates that the LW strength does not significantly affect the primordial IMF. As the range of background LW field strengths tested here covers the most likely values from literature, we infer that the IMF for Pop III.2 stars is not significantly different from the initial population of Pop III.1 stars.

We also show the IMF from \citetalias{Prole2022} as a dotted line, which was produced from idealised halos as opposed the cosmological initial conditions. The cosmological halos produced a bi-modal distribution with a significant population of brown dwarfs that the idealised halos are missing. Even when ignoring the brown dwarfs, the cosmological initial conditions have yielded distributions tending to lower mass pre-stellar cores.

\begin{figure}
	 \hbox{\hspace{-0.5cm} \includegraphics[scale=0.7]{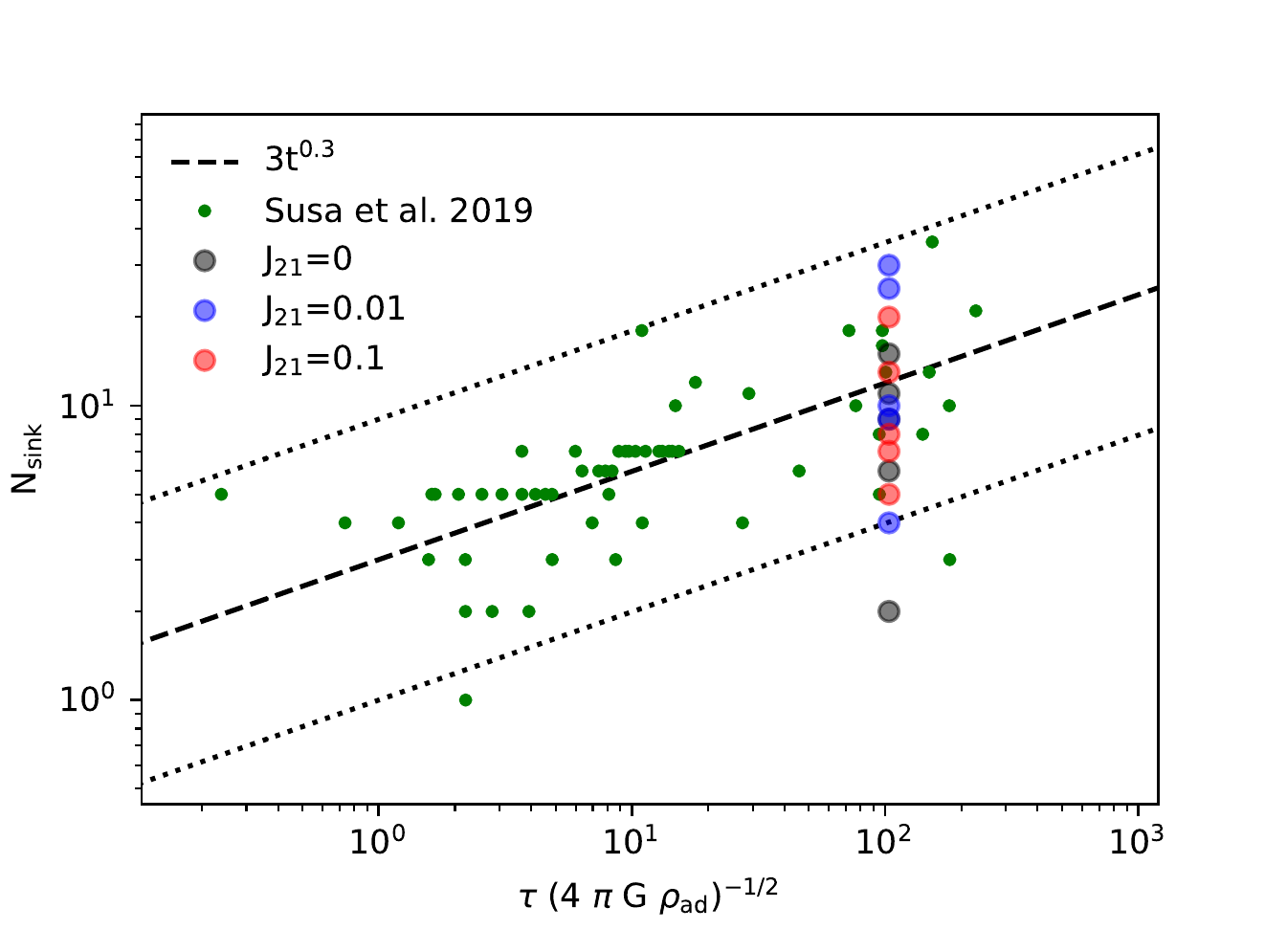}}
    \caption{Comparison of the results of our study with the data of various studies collected in Figure 10 of \citet{Susa2019}. The number of sink particles formed is plotted verses time scaled as $\tau / \sqrt{4 \pi G \rho_{\rm ad}}$ . The thick dashed line shows the fit $\propto t^{0.3}$ and the dotted lines denote x3 and x1/3 of the thick dashed line. }
    \label{fig:susa}
\end{figure}

To compare the fragmentation behaviour of our simulations with the results of previous studies, we have recreated Figure 10 of \cite{Susa2019} in Figure \ref{fig:susa}, which collected data from various studies. They found a power-law growth of the number of fragments with time. As the studies implemented different maximum densities, the time at the end of the simulations was scaled as $\tau / \sqrt{4 \pi G \rho_{\rm ad}}$, where $\rho_{\rm ad}=1.67 \times 10^{-5}$ is the maximum density of \cite{Greif2012} (see \citealt{Susa2019} for further details). The fragmentation behaviour in our simulations agrees well with the findings of \cite{Susa2019}. Our results appear advanced in time compared to the other studies despite only following 300 yr of fragmentation because of the high maximum density implemented in this study.

\section{Caveats} 
\label{sec:caveats}
Aside from the obvious uncertainties in the $\Lambda$CDM model on which this work and the work of \citetalias{Schauer2021} are based on, there are a number of caveats to note.

The LW fields we have used in this study assume a population of massive Pop III stars already exists. While previous studies do suggest that Pop III stars were massive, more recent work suggests that they may only grow to a few M$_\odot$ (e.g. \citealt{Stacy2013,Wollenberg2020,Prole2022,Jaura2022}). In this scenario, significantly less LW radiation would be produced, however radiation from solar mass stars can inhibit H$_2$ formation through the destruction of H$^-$. The fields used in this study therefore represent the maximum effect Pop III stars can have on the the Pop III.2 stars that follow. Since the effects of the LW fields appear to be insignificant, it is likely that this result also represents the outcome for weaker fields.

The average halo mass from the $J_{21}=0$ simulations of  \citetalias{Schauer2021} are a few times larger than that of \cite{Hirano2015}. We assume this is due to the difference in self-shielding treatment, as \cite{Hirano2015} uses the \cite{Wolcott-Green2011} method. For example, an overestimation of self-shielding would result in an underestimation of photodissociation and hence a lower average mass of star forming halos. However, here we have shown that the IMF above the brown dwarf limit is invariant to halo mass.

We have assumed that the pre-stellar core radius grows as Equation \ref{eq:radius}. This process begins immediately after sink particle formation. While this is predicted by stellar theory, it is unclear at what point the pre-stellar core would begin to expand, which may affect the accretion behaviour.

We have neglected the presence of primordial magnetic fields in this study. While the findings of \cite{Prole2022a} suggest that primordial magnetic fields make little difference to the primordial IMF due to their small-scale structure, the field is still active, with other recent studies finding that magnetic fields indeed lead to higher mass Pop III stars (e.g. \citealt{Saad2022,Hirano2022,Stacy2022}).

\section{Conclusions}
The results of cosmological simulations by \citetalias{Schauer2021} show that increasing the background LW field strength increases the average halo mass required for star formation. These simulations ran up to a maximum density of $10^{-19}$ g cm$^{-3}$, hence the knock-on effects on the Pop III IMF were unclear. In this investigation, we have performed follow-up simulations of 5 halos for each of the $J_{21}$ = 0, 0.01 and 0.1 LW field strengths, resolving the pre-stellar core density of $10^{-6}$ g cm$^{-3}$ before inserting sink particles and following the fragmentation behaviour for hundreds of years further. We have found that the mass accreted onto sinks by the end of the simulations is proportional to the mass within the $\sim 10^{-2}$ pc molecular core, which is not correlated to the initial mass of the halo. As such, the IMF shows little dependence on the LW field strength. We also find no clear relationship between the estimated accretion time of gas lying further out within the halo and the LW field strength, suggesting that the LW field is unlikely to influence the development of the IMF at later times. As the range of background LW field strengths tested here covers the most likely values from literature, we conclude that the IMF for so-called Pop III.2 stars is not significantly different from that of the initial population of Pop III.1 stars, although we cannot rule out greater effects in the small subset of halos that are illuminated by LW fields much stronger than the average value \citep[see e.g][]{Latif2014}. In a future paper, we will explore the effects of increasing the streaming velocities between the gas and dark matter on the Pop III IMF, as this has been shown to increase halo masses through a different mechanism.

\label{sec:conclusion}

\section*{Acknowledgements}
This work used the DiRAC@Durham facility managed by the Institute for Computational Cosmology on behalf of the STFC DiRAC HPC Facility (www.dirac.ac.uk). The equipment was funded by BEIS capital funding via STFC capital grants ST/P002293/1, ST/R002371/1 and ST/S002502/1, Durham University and STFC operations grant ST/R000832/1. DiRAC is part of the National e-Infrastructure.

The authors gratefully acknowledge the Gauss Centre for Supercomputing e.V. (www.gauss-centre.eu) for supporting this project by providing computing time on the GCS Supercomputer SuperMUC at Leibniz Supercomputing Centre (www.lrz.de) under project pr53ka. AS was partially supported by NSF grant AST-1752913.

RSK and SCOG acknowledge computing resources provided by the Ministry of Science, Research and the Arts (MWK) of the State of Baden-W\"{u}rttemberg through bwHPC and the German Research Foundation (DFG) through grant INST 35/1134-1 FUGG and for data storage at SDS@hd through grant INST 35/1314-1 FUGG. 

RSK and SCOG acknowledge financial support from DFG via the collaborative research center (SFB 881, Project-ID 138713538) “The Milky Way System” (subprojects A1, B1, B2 and B8), from the Heidelberg Cluster of Excellence “STRUCTURES” in the framework of Germany’s Excellence Strategy (grant EXC-2181/1, Project-ID 390900948) and from the European Research Council (ERC) via the ERC Synergy Grant “ECOGAL” (grant 855130). RSK furthermore thanks the German Ministry for Economic Affairs and Climate Action for funding in the project ``MAINN'' (funding ID 50OO2206).

We also acknowledge the support of the Supercomputing Wales project, which is part-funded by the European Regional Development Fund (ERDF) via Welsh Government.


\section*{Data Availability}
The data underlying this article will be shared on reasonable request to the corresponding author.



\bibliographystyle{mnras}

\bibliography{references.bib} 




\bsp	
\label{lastpage}

\end{document}